Title: An *in situ* comparison of electron acceleration at collisionless shocks under differing upstream magnetic field orientations

Short title: Examining shock-acceleration of electrons

A. Masters[1], A. H. Sulaiman[2], Ł. Stawarz[3], B. Reville[4], N. Sergis[5,6], M. Fujimoto[7,8], D. Burgess[9], A. J. Coates[10,11], M. K. Dougherty[1].

[1]The Blackett Laboratory, Imperial College London, Prince Consort Road, London, SW7 2AZ, UK.
[2]Department of Physics and Astronomy, University of Iowa, Iowa City, IA, USA 52242.
[3]Astronomical Observatory, Jagiellonian University, ul. Orla 171, 30-244 Krakow, Poland.
[4]School of Mathematics and Physics, Queens University Belfast, Belfast BT7 1NN, UK.
[5]Office of Space Research and Technology, Academy of Athens, Soranou Efesiou 4, 11527 Athens, Greece.
[6]Institute of Astronomy, Astrophysics, Space Applications and Remote Sensing, National Observatory of Athens, Athens, Greece.
[7]Institute of Space and Astronautical Science, Japan Aerospace Exploration Agency, 3-1-1 Yoshinodai, Chuo-ku, Sagamihara, Kanagawa 252-5210, Japan.
[8]Earth-Life Science Institute, Tokyo Institute of Technology, 2-12-1 Ookayama, Meguro, Tokyo 152-8551, Japan.
[9]School of Physics and Astronomy, Queen Mary University of London, London E1 4NS, UK.
[10]Mullard Space Science Laboratory, Department of Space and Climate Physics, University College London, Holmbury St. Mary, Dorking RH5 6NT, UK.
[11]The Centre for Planetary Sciences at UCL/Birkbeck, Gower Street, London WC1E 6BT, UK.
Corresponding author: A. Masters (a.masters@imperial.ac.uk)

Abstract

A leading explanation for the origin of Galactic cosmic rays is acceleration at high-Mach number shock waves in the collisionless plasma surrounding young supernova remnants. Evidence for this is provided by multi-wavelength non-thermal emission thought to be associated with ultrarelativistic electrons at these shocks. However, the dependence of the electron acceleration process on the orientation of the upstream magnetic field with respect to the local normal to the shock front (quasi-parallel/quasi-perpendicular) is debated. Cassini spacecraft observations at Saturn's bow shock has revealed examples of electron acceleration under quasi-perpendicular conditions, and the first *in situ* evidence of electron acceleration at a quasi-parallel shock. Here we use Cassini data to make the first comparison between energy spectra of locally accelerated electrons under these differing upstream magnetic field regimes. We present data taken during a quasi-perpendicular shock crossing on 2008 March 8 and during a quasi-parallel shock crossing on 2007 February 3, highlighting that both were associated with electron acceleration to at least MeV energies. The magnetic signature of the quasi-perpendicular crossing has a relatively sharp upstream-downstream transition, and energetic electrons were detected close to the transition and immediately downstream. The magnetic transition at the quasi-parallel crossing is less clear, energetic electrons were encountered upstream and downstream, and the electron energy spectrum is harder above ~100 keV. We discuss whether the acceleration is consistent with diffusive shock acceleration theory in each case, and suggest that the quasi-parallel spectral break is due to an energy-dependent interaction between the electrons and short, large-amplitude magnetic structures.

1. Introduction

Collisionless shock waves are ubiquitous in the highly tenuous space plasma environments of our Solar System, as well as in a diverse range of similarly collisionless astrophysical plasma environments (see the review by Treumann 2009). Like shocks in collisional neutral fluids, collisionless plasma shocks also form when the speed of flow relative to an obstacle is greater than the speed at which information can be transferred via the medium. However, at collisionless plasma shocks the dissipation of energy is achieved by interactions between the charged particles and the electromagnetic field, and the relevant information transfer speed is the speed of fast magnetosonic waves, whereas it is the speed of sound waves in the case of collisional neutral fluids.

Key parameters used to describe collisionless shocks include the shock Mach numbers and the shock angle. A Mach number is defined in the shock rest frame as the upstream flow speed normal to the shock surface divided by an upstream wave speed. The fast magnetosonic Mach number ($M_f$) and the Alfvén Mach number ($M_A$) are related to the upstream speed of fast magnetosonic and Alfvén waves respectively, where the fast magnetosonic Mach number indicates how much bulk flow kinetic energy must be dissipated at the shock. The shock angle ($\theta_{Bn}$) is the angle between the upstream magnetic field and the local normal to the shock surface. Changing this parameter has a significant impact on the physics of the shock, since the upstream field orientation strongly influences particle trajectories, including the motion of suprathermal particles across the shock front. Typically, shocks with $\theta_{Bn} < 45°$ are referred to as quasi-parallel, whereas those with $\theta_{Bn} > 45°$ are referred to as quasi-perpendicular.

Much of the drive to understand how collisionless shocks work is motivated by the historic cosmic ray problem. The leading mechanism for producing cosmic rays at energies from ~$10^{10}$ eV up to ~$10^{15}$ eV is acceleration at the shock waves that surround young (<1000 year-old) supernova remnants (SNRs; e.g., Blandford & Eichler 1987). The sub-relativistic collisionless shocks surrounding these SNRs are very high Mach number, and the process thought to accelerate a fraction of the thermal pool particles to very high energies with high overall efficiency is known as Diffusive Shock Acceleration (DSA; e.g., Bell 1978a, 1978b; Drury 1983; Blandford & Eichler 1987; Jones & Ellison 1991). DSA is a first-order Fermi process where particles bounce between scattering centers located both upstream and downstream of the shock front, gaining energy in the process due to the convergence of these scattering centers that results from bulk flow deceleration across the shock. The proposed scattering centers are magnetohydrodynamic (MHD) fluctuations.

Electron acceleration at young SNR shocks is of particular interest, since remote evidence supporting the operation of DSA at these shocks is provided by the detection of radio, x-ray, and gamma-ray non-thermal emission associated with ultrarelativistic electrons (Aharonian et al. 2004; Uchiyama et al. 2007; Reynolds 2008; Abdo et al. 2011; Helder et al. 2012). The process by which thermal electrons are accelerated to energies at which they interact with MHD-scale fluctuations has been debated (the so-called electron "injection" problem), as well as how the subsequent DSA of electrons is influenced by local conditions, particularly the shock angle (Jokipii 1987). These debates have often been centered on the remnant of SN1006 (e.g., Koyama et al. 1995), where the regions of more and less intense x-ray emission surrounding the remnant are thought to result from differing local upstream magnetic field orientations (e.g., Bocchino et al. 2011).

*In situ* observations of collisionless shocks in the solar wind have allowed significant progress in this field (see the reviews by Russell 1985; Smith 1985; Burgess 2007); however, these shocks are generally far lower Mach number than those that surround young SNRs. Recently, data taken by the Cassini spacecraft at the bow shock wave that stands in the solar wind in front of Saturn have been analyzed. Due to increasing solar wind Mach numbers with distance from the Sun (e.g., Slavin & Holzer 1981), this shock wave is one of the highest Mach number shocks ever observed *in situ*, occasionally bridging the gap to the young SNR Alfvén Mach number regime (Achilleos et al. 2006; Masters et al. 2011). Studies based on Cassini observations have provided the first evidence for shock reformation at high Mach numbers (Sulaiman et al. 2015, 2016). Furthermore, evidence that electron "injection" occurs at all shock angles at sufficiently high Mach numbers (like those of young SNR shocks) has been provided by a study that reported the first evidence for electron acceleration at a quasi-parallel shock (Masters et al. 2013), consistent with numerical modeling work (Guo & Giacalone 2015). A later study that examined electron acceleration at hundreds of Cassini shock crossings also supports this conclusion (Masters et al. 2016).

In this paper we analyze two Cassini crossings of Saturn's bow shock where rare evidence for electron acceleration to relativistic (~MeV) energies has been identified. One of these is quasi-perpendicular and the other is quasi-parallel, which provides us with our first opportunity to make an *in situ* comparison of electron acceleration at shocks under differing upstream magnetic field orientations, highly relevant for the problem of particle acceleration at young SNRs.

2. Observations

The Cassini spacecraft has been in Saturn orbit since July 2004. During its orbital tour the spacecraft has regularly sampled the near-Saturn solar wind, resulting in hundreds of crossings of Saturn's bow shock. These crossings took place predominantly on the dayside of the shock surface, and under a range of upstream conditions.

Data taken during shock crossings by Cassini have been surveyed by Sulaiman et al. (2016) and Masters et al. (2016), who discuss the extent of information about each crossing that can be extracted from Cassini data sets. These studies provide an estimated Alfvén Mach number of each crossing, as well as a separation of the crossings by shock geometry (quasi-parallel/quasi-perpendicular). The typical Alfvén Mach number of the Cassini shock crossings is ~15, with instances of lower (~5) and higher (~100) values. The crossings are generally quasi-perpendicular, due to the prevailing direction of the (variable) interplanetary magnetic field at Saturn's heliocentric distance.

This study follows directly from the results presented by Masters et al. (2016), who searched for evidence of electron acceleration at these Cassini shock crossings. These authors identified three crossings with particularly strong energetic electron signatures, which cannot be explained as a result of leakage of energetic electrons from within Saturn's magnetic field environment. Of these three most striking examples with shock-accelerated electrons, two of the crossings are quasi-perpendicular and at typical Alfvén Mach numbers, one inbound (upstream-downstream) and one outbound (downstream-upstream). The other example is the inbound, high-Alfvén Mach number quasi-parallel crossing previously reported by Masters et al. (2013). Since the present study concerns the influence of upstream conditions, we focus on the inbound quasi-perpendicular and inbound quasi-parallel crossings only. Note that the excluded (outbound) quasi-perpendicular crossing has a similar signature to the included (inbound) quasi-perpendicular crossing (i.e., inclusion of the outbound crossing has no impact on the conclusions drawn here). We refer the reader to Masters et al. (2016) for a detailed discussion of electron acceleration signatures at Saturn's bow shock, including an explanation of why only a few strong signatures have been observed.

Data taken by three instruments mounted on the three-axis-stabilized Cassini spacecraft during the two shock crossings of interest are presented here. The Cassini magnetometer measures the local magnetic field vector (Dougherty et al. 2004). The Electron Spectrometer (ELS) and Ion Mass Spectrometer (IMS) of the Cassini Plasma Spectrometer (Young et al. 2004) detect electrons in the 0.5 eV to 26 keV energy range, and ions with energy-per-charge between 1 V and 50 kV, respectively. The Low Energy Measurements System (LEMMS) of the Magnetospheric Imaging Instrument (Krimigis et al. 2004) detects electrons in the 18 keV to ~1 MeV energy range. All particle detectors have a limited Field-Of-View (FOV).

Figure 1 shows two hours of data taken by Cassini on 2008 March 8, encompassing the selected quasi-perpendicular shock encounter. The magnetic field magnitude signature shown in Figure 1a reveals a clear upstream to downstream transition at ~21:18 Universal Time (UT), and a higher level of magnetic field variability in the downstream region than in the upstream region. Figure 1b shows the shock angle as a time series, computed for each magnetic field vector using a normal to the local shock surface predicted by a global shape model (Went at al. 2011). Throughout the interval considered the nearby shock surface is expected to have been quasi-perpendicular. Figure 1c quantifies the level of magnetic field variability. The parameter presented, $\delta B / <B>$, has been calculated based on a 300 s window (10 times the typical timescale of the downstream magnetic field fluctuations), which was centered on each data point to give an associated value of the background field strength, $<B>$, as the median of all data points within the window. The quantity $\delta B$ is then defined as each field strength measurement minus the corresponding background field strength. The parameter $\delta B / <B>$ is therefore essentially a measure of the field strength fluctuations normalized to the background value. This approach highlights the low level of upstream magnetic field variability in comparison to the field fluctuations near the shock front and downstream.

Figures 1d and 1e show energy-time spectrograms of electron differential intensity over the energy range 0.5 eV to ~1 MeV, combining data taken by ELS and LEMMS. The upstream-downstream transition is clear in the thermal electrons (Figure 1e), also occurring at ~21:18 UT. The electrons detected upstream at energies below 100 eV are a superposition of the ambient solar wind electron population and a population of spacecraft photoelectrons, whereas downstream the heated ambient population is better distinguished from the lower energy (<10 eV) spacecraft photoelectrons. Note that the modulation at a period of ~7 minutes is related to changes in the ELS sensor FOV during instrument actuation. The signature of more energetic, shock-accelerated electrons (Figure 1d) was observed immediately before (from ~21:16 UT), during, and after the approximate time of the thermal electron transition (~21:18 UT), with progressively lower intensities in all LEMMS energy channels with increasing time in the downstream region. Finally, Figure 1f shows an energy-time spectrogram of thermal ion count rate measured by IMS, which also reveals the shock transition at ~21:18 UT. Before the transition the upstream (antisunward) plasma flow direction was not within the IMS FOV, although a population of ~10 keV ions was detected at ~21:15 UT, most likely a signature of the incident solar wind ions that had been reflected back upstream at the shock front. In the downstream region the population of heated solar wind ions was regularly resolved (~7-minute periodic modulation also related to changes in the sensor FOV during instrument actuation).

We refer the reader to Masters et al. (2011, 2016) and Sulaiman et al. (2016) for a discussion of the information that can be reliably extracted from Cassini data taken at Saturn's bow shock. The location of the shock crossing shown in Figure 1 and the mean upstream magnetic field strength of ~0.8 nT indicates an Alfvén Mach number of $M_A$~15 (assuming a stationary shock in the planetary rest frame; see Sulaiman et al., 2016). Combining with typical upstream plasma parameters at Saturn orbit (e.g., upstream flow speed ~450 km s$^{-1}$; e.g., Slavin & Holzer 1981) this corresponds to a fast magnetosonic Mach number of $M_f$~10. Note that the large relative uncertainty in the motion of the shock surface throughout the two-hour interval prevents a reliable transformation from temporal to spatial coordinates (e.g., Masters et al. 2011).

Figure 2 shows two hours of data taken by Cassini on 2007 February 3, encompassing the selected quasi-parallel shock encounter, in the same format as that of Figure 1. This event was reported by Masters et al. (2013), and to the best of our knowledge remains the only confirmed case of *in situ* evidence for acceleration of electrons at a quasi-parallel shock. Figures 2a through 2c reveal a more extended upstream to downstream magnetic transition in this quasi-parallel case, characterized by a high-level of magnetic field variability both upstream and downstream. At ~00:05 UT the upstream magnetic field direction changed to produce a locally quasi-parallel shock from that time until beyond the end of the interval shown. The high level of upstream magnetic field variability that is typical of quasi-parallel shocks produced a highly variable shock angle based on the magnetic field vector time series, shown in Figure 2b. However, taking the average magnetic field vector in the interval 00:10 to 01:00 UT gives an expected low shock angle of ~20°. Note that in Figure 2c a window duration of 200 s has been used (10 times the typical timescale of the downstream magnetic field fluctuations).

Figures 2d and 2e also show clear differences between this quasi-parallel case and the previous quasi-perpendicular case. The time of the shock transition is most clear in the thermal electron signature (Figure 2e, occurring at ~01:05 UT), although the low-energy electron distribution is more variable. The signature of energetic, shock-accelerated electrons (Figure 2d) peaks at approximately the transition time, similar to the quasi-perpendicular case. However, in contrast to the quasi-perpendicular case the signature of electron acceleration at this quasi-parallel shock begins well before the thermal plasma transition (at ~01:05 UT), first resolved at ~00:35 UT. The IMS data shown in Figure 2f reveals upstream features that we identify as a "diffuse" ion population that is typical of quasi-parallel shocks.

The weak upstream field strength (~0.1 nT) at this quasi-parallel shock encounter resulted in an unusually high Alfvén Mach number of $M_A$~100, approaching the high-Mach number regime of young SNR shocks. The corresponding fast magnetosonic Mach number was also relatively high, $M_f$~25 (Achilleos et al. 2006; Masters et al. 2013).

Figure 3 compares the electron energy spectra measured by ELS and LEMMS at the quasi-perpendicular and quasi-parallel crossings. In both Figures 3a and 3b the spectra have been averaged over a two-minute interval when the signature of accelerated electrons was strongest (highest LEMMS electron channel intensities). In both cases a non-thermal population that extends from the thermal population to higher energies is present in the ELS data (Masters et al. 2016). The observations indicate a transition from a harder to a softer electron energy spectrum with increasing energy between ~5 and ~18 keV, particularly in the quasi-parallel case, although comparison between ELS and LEMMS spectra should be treated with caution due to the problem of inter-calibration between the sensors. Another similarity between the quasi-perpendicular and quasi-parallel spectra are the absolute differential intensities of suprathermal electrons detected by LEMMS (>18 keV). Power-law fits to this higher energy range of the electron energy spectrum have been made, and are shown in Figure 3. These describe straight lines on such log-log scales, and the slope (gradient) of each line given in Figure 3 is the index of the associated best-fit power law. The uncertainty in each slope in quoted to one significant figure and dictates the accuracy to which the best-fit slope is given.

The suprathermal electrons at the quasi-perpendicular shock can be described by a single power law distribution (Figure 3a). As indicated earlier in this section, at the excluded outbound quasi-perpendicular crossing that also has a strong energetic electron signature the LEMMS observations show similar channel intensities, the spectrum can also be described by a single power law, and the calculated slope is within uncertainties of the value associated with the presented inbound quasi-perpendicular shock (Figure 3a). In contrast, at the quasi-parallel shock this energetic population cannot be described by a single power law to within the measurement uncertainties (Figure 3b). Instead, two separate power laws are consistent with the quasi-parallel shock observations, where a transition from a softer to a harder spectrum with increasing energy occurs at ~100 keV. This transition at ~100 keV is identifiable throughout the LEMMS electron observations made between ~00:30 and ~01:20 UT on 2007 February 3, which is approximately the entire interval during which the signature of shock-accelerated electrons was resolved at the quasi-parallel shock. Note that the index associated with a power-law fit to the electron energy spectrum above 100 keV is the same at the quasi-perpendicular and quasi-parallel shocks, to within the uncertainties.

3. Discussion

This comparison between electron acceleration at a quasi-perpendicular and at a quasi-parallel shock reveals clear differences, both in the magnetic structure of the shock and in the signature of suprathermal electrons. However, electron intensities measured at the highest energies (~20 keV to ~1 MeV) are comparable, showing that regardless of the highlighted differences both quasi-perpendicular and quasi-parallel shocks are capable of accelerating thermal electrons to relativistic energies with analogous overall efficiencies.

Below we discuss what these Cassini observations tell us about the electron acceleration process under each upstream magnetic field orientation. In addition to drawing on published theories and numerical modeling results we also compare with predictions of the DSA theory (e.g., Bell 1978a). DSA theory for high-Mach number shocks (ratio of downstream to upstream plasma density equal to 4) predicts that in the test-particle limit the momentum distribution of isotropic, shock-accelerated particles, $f(p)$, is described by a universal power-law spectrum, where $f(p) \sim p^{-4}$. Relating this to differential intensity (shown in Figures 1 through 3), this intensity, $I$, is predicted to be described as $I \sim p^2 f(p)$ (e.g., Forman 1970). Hence, the theory predicts that the plots of electron differential intensity against electron kinetic energy, $E$, shown in Figure 3 should show a power-law relationship, where $I \sim E^{-1}$. The comparison between data and this prediction is appropriate in the LEMMS energy range (non-thermal, >18 keV), where observations are consistent with power laws. Note that due to the upper limit of the LEMMS energy range we do not expect to be able to resolve a trans-relativistic effect at higher energies (a transition to an $E^{-2}$ dependence at ~MeV energies).

Both the magnetic and accelerated electron signatures of the quasi-perpendicular shock crossing shown in Figure 1 are typical of extensive past observations made at shocks in the heliosphere (e.g., Oka et al. 2006). The "injection" and subsequent acceleration of thermal electrons at a quasi-perpendicular shock has been the subject of much discussion in the literature, where shock drift acceleration, growth of the Buneman instability and its influence on shock surfing acceleration, the impact of nonstationarity, the role of ion-scale shock surface fluctuations, and the frequently invoked role of whistler waves have all been studied (Levinson 1992; Burgess 2006; Amano & Hoshino 2007, 2009a, 2009b, 2010; Riquelme & Spitkovsky 2011; Matsukiyo & Scholer 2012; Matsumoto et al. 2012, 2015). The apparent "injection" of thermal electrons close to the shock front is consistent with these ideas, for example, the mechanism described by Amano & Hoshino (2010) where thermal electrons undergo shock drift acceleration and are then scattered by self-generated whistler waves. This is thought to be possible at all Alfvén Mach numbers for perpendicular shocks, and is highlighted here because of its associated prediction for quasi-parallel shocks, discussed below in the context of our quasi-parallel crossing.

The differential intensity of energetic electrons at the quasi-perpendicular shock shown in Figure 3a (>18 keV) can be described by a power law with an index of -2.5±0.5. This energy spectrum is therefore softer than the DSA theory prediction of an index of -1. However, although this may hint that the DSA model, at least in its simplest version discussed here, is not fully applicable, the fact that the spatial scale of the region in front of the shock where shock-reflected ions are present is comparable to the gyroradius of an MeV electron (both ~2000 km; Gosling & Thomsen 1985) is consistent with an electron acceleration process that could nonetheless be described as "diffusive shock acceleration".

Although the shock encounter shown in Figure 2 is presently the only reported example of electron acceleration under quasi-parallel upstream conditions, the magnetic structure of the shock is typical of past observations of quasi-parallel shocks in general. The presence of counter-streaming ion populations upstream of quasi-parallel shocks is known to lead to significant local enhancements of the magnetic field, often referred to as Short Large-Amplitude Magnetic Structures (SLAMS, Schwartz & Burgess 1991; Schwartz et al. 1992), and such structures are indeed evident in Figures 2a and 2c. Note that largely due to their presence of these SLAMS the local conditions at the shock front, and in particular the local shock angle, are highly variable (see Figure 2b).

What is atypical for this quasi-parallel shock, however, in addition to a particularly strong signature of energetic electrons, is the low (~0.1 nT) upstream magnetic field strength and the resulting high Alfvén Mach number (~100), and this suggests a fundamental link between the efficiency of the electron "injection" process and the shock Mach number. The mechanism outlined by Amano & Hoshino (2010), mentioned above in the context of our quasi-perpendicular shock crossing, makes the prediction that thermal electron "injection" takes place in the near upstream, and is only possible at sufficiently high Mach numbers for quasi-parallel shocks. This is consistent with the observations presented here, where the acceleration of electrons directly from the thermal pool has been identified in the region where the main thermal plasma transition occurred (Masters et al. 2013). In addition, recent studies have proposed that electron "injection" can occur locally farther upstream, possibly associated with foreshock phenomena (Wilson et al. 2016).

Without knowledge of the shock location with respect to the spacecraft we are unable to transform Figure 2 from a temporal to a spatial scale. However, an approximation of the scale of the upstream region during which shock-accelerated electrons were resolved (~00:30 to ~01:00 UT) is ~200,000 km, based on shock motion toward the spacecraft at 100 km s$^{-1}$ (Achilleos et al. 2006; Masters et al. 2011). This is of order 10 times the gyroradius of an electron at an energy of 1 MeV in the upstream magnetic field of ~0.1 nT.

The hardening of the accelerated electron energy spectrum at ~100 keV at the quasi-parallel shock (Figure 3b) suggests two distinct regimes in the electron acceleration process. A power-law fit to the differential intensity below 100 keV returns a power-law index of -4±1, which is a softer electron energy spectrum than in the quasi-perpendicular case (in the same energy range). In contrast, a power-law fit above 100 keV gives an index of -2±1, which is closer to the DSA theory prediction of -1. In the following discussion we explore the potential role of the SLAMS identified throughout the shock crossing (Figure 2c) in creating this electron energy spectral break.

Figure 4 shows magnetic field data taken during a sub-interval of the interval shown in Figure 2. This period contains two SLAMS identified upstream of this quasi-parallel shock. The results of minimum variance analysis (Sonnerup & Scheible 1998) applied to data taken during the spacecraft encounter with the first structure reveals a right-handed, approximately elliptical polarization about the background magnetic field direction in the spacecraft frame. The polarization of the second structure is less clear, potentially due to the less well-constrained minimum variance direction of the field. Past observations of SLAMS at Earth's bow shock suggest that they grow directly from the upstream wave field and attempt to propagate away from the shock, but are advected towards the shock with the bulk plasma flow (e.g., Schwartz et al. 1992). These structures are therefore expected to be intrinsically left-hand polarized, with an apparent right-handed polarization in the spacecraft frame. Taking the typical timescale of the SLAMS in Figure 1 (~20 s) and multiplying by the upstream flow speed (~500 km s$^{-1}$, Masters et al. 2013) gives the spatial scale of these structures as ~10,000 km. In the upstream magnetic field of 0.1 nT the energy at which the electron gyroradius equals the above SLAMS spatial scale is ~90 keV.

We therefore propose that the break in the quasi-parallel shock electron energy spectrum at ~100 keV is due to an energy-dependent interaction between the electrons and SLAMS. Kuramitsu & Hada (2008), in examining the transport of charged particles in an idealized SLAMS have previously found the transition from adiabatic to non-adiabatic behavior to occur at an energy corresponding to near spatial resonance between the gyroradius of the charged particles and the SLAMS. Thus for the conditions described above, electrons with energies less then 100 keV can be trapped in these non-linear structures and effectively swept out with the flow. This adiabatic trapping will be concomitant with a significant increase in the particle anisotropy which, when averaged over long timescales, naturally leads to deviations from the standard DSA theory, and a softening of the spectrum. In contrast, higher-energy (above ~100 keV) electrons with a gyroradius exceeding the scale of the SLAMS cannot be trapped inside such non-linear magnetic structures and in fact may have an enhanced isotropization rate due to scattering on multiple SLAMSs, thus participating more efficiently in the diffusive shock acceleration process. The break in the energy spectrum at ~100 keV may therefore mark the transition to genuinely diffusive behavior. Revealing the details of the physics controlling this identified feature of the energy spectrum of electrons accelerated at the quasi-parallel shock requires further work.

Although the limited spatial scale and highly variable upstream conditions at Saturn's bow shock place some limits on the extent to which we can draw conclusions about electron acceleration at young supernova remnant shocks, this examination of the influence of the upstream magnetic field orientation nonetheless suggests that quasi-perpendicular and quasi-parallel shocks are similarly effective electron accelerators at high Mach numbers. We also note that magnetic field amplification via the non-resonant current instability, while not thought to be occurring in the cases presented here, is expected to lead to a similarly highly structured magnetic field (Bell 2004, Reville & Bell, 2013), which may provide an alternative method for producing deviations from pure power-law measurements in synchrotron observations of supernova remnants.

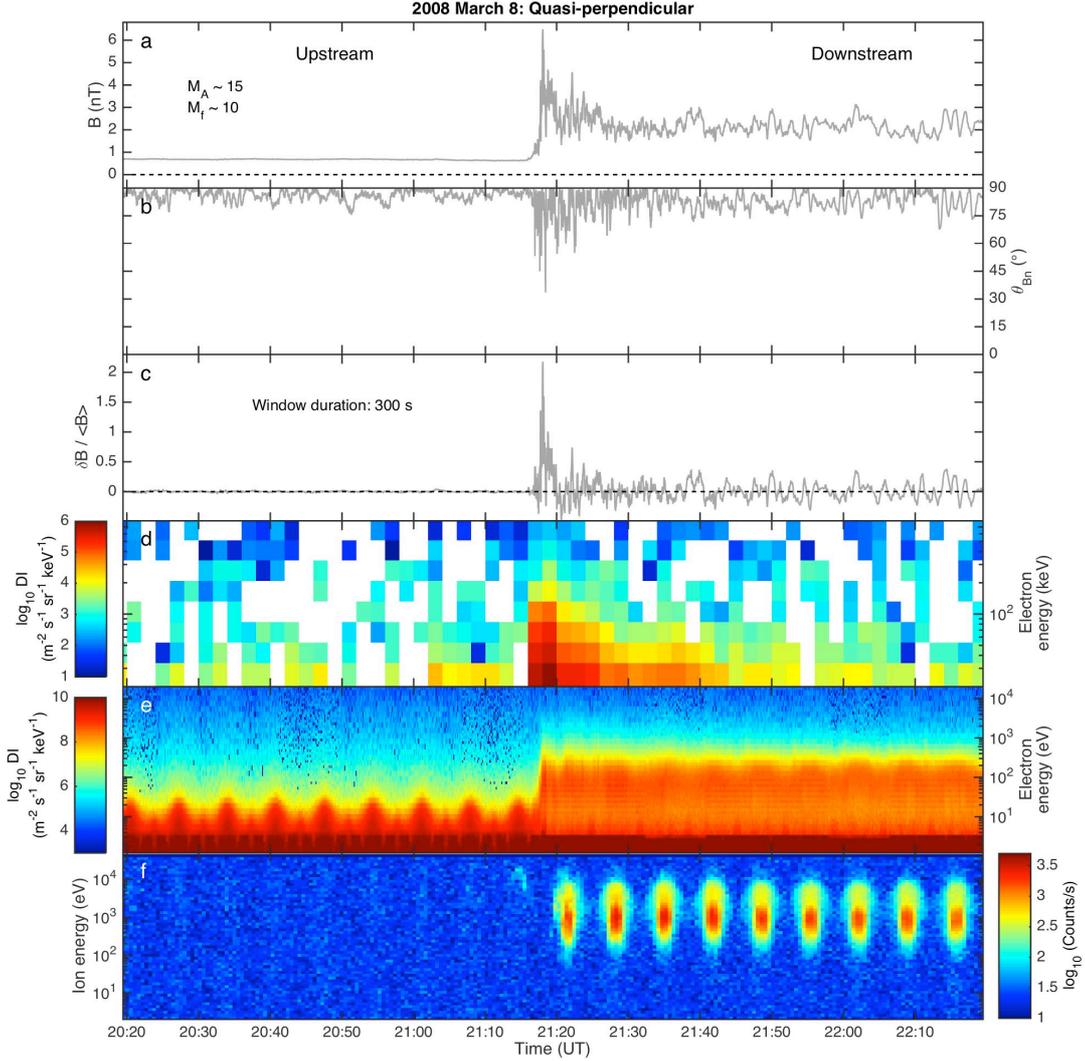

Figure 1. *In situ* observations made by Cassini on 2008 March 8 over a two-hour interval encompassing an inbound crossing of Saturn's bow shock under quasi-perpendicular upstream magnetic conditions. (a) Magnetic field magnitude. (b) Shock angle ($\theta_{Bn}$, see Section 2). (c) Normalized magnitude of magnetic field fluctuations ($\delta B/\langle B \rangle$, see section 2, window duration ~10 times the typical timescale of downstream magnetic field fluctuations). (d) Energy-time spectrogram of electron Differential Intensity (DI) at energies above 18 keV (LEMMS). (e) Energy-time spectrogram of electron Differential Intensity (DI) at energies below 18 keV (ELS anode 5). (e) Energy-time spectrogram of ion count rate (IMS anode 5).

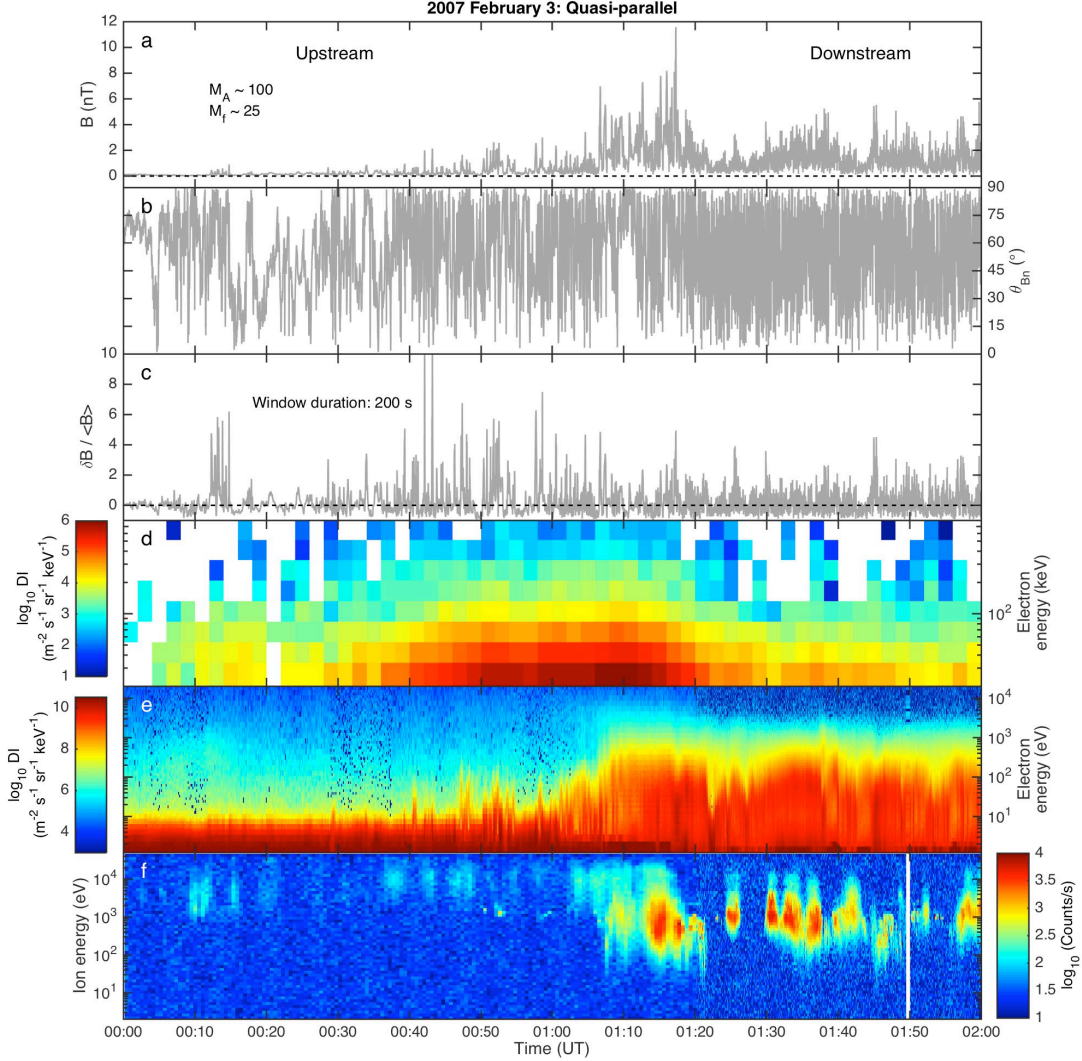

Figure 2. *In situ* observations made by Cassini on 2007 February 3 over a two-hour interval encompassing an inbound crossing of Saturn's bow shock under quasi-parallel upstream magnetic conditions. (a) Magnetic field magnitude. (b) Shock angle ($\theta_{Bn}$, see Section 2, window duration ~10 times the typical timescale of magnetic field fluctuations). (c) Magnitude of magnetic field fluctuations ($\delta B/<B>$, see section 2). (d) Energy-time spectrogram of electron Differential Intensity (DI) at energies above 18 keV (LEMMS). (e) Energy-time spectrogram of electron Differential Intensity (DI) at energies below 18 keV (ELS anode 5). (e) Energy-time spectrogram of ion count rate (IMS anode 5).

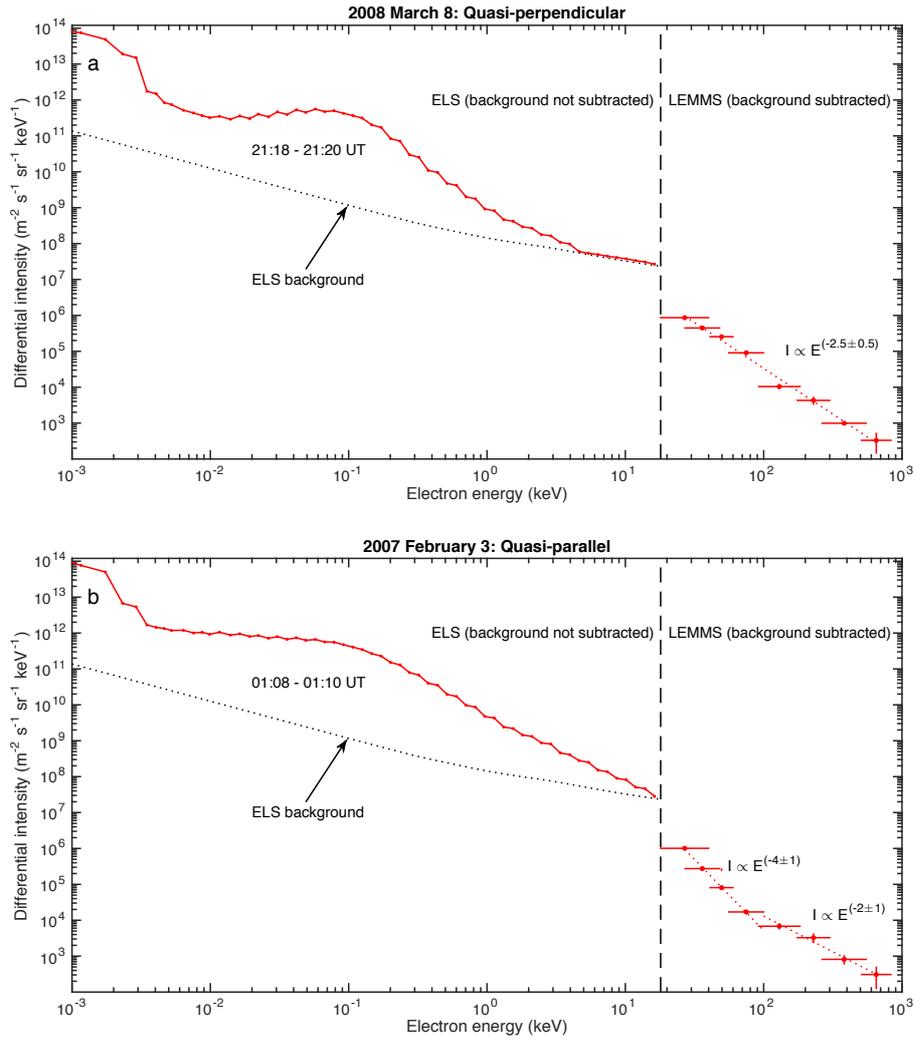

Figure 3. Comparison of electron energy spectra measured at quasi-perpendicular and quasi-parallel shock encounters where local electron acceleration to ~MeV energies took place. A combination of ELS (<18 keV) and LEMMS (>18 keV) data are shown. (a) Two-minute-averaged spectrum at the 2008 March 8 quasi-perpendicular shock crossing. (a) Two-minute-averaged spectrum at the 2007 February 3 quasi-parallel shock crossing. "Step-like" features in the ELS energy range are due to onboard spacecraft averaging in response to telemetry constraints. Dotted red lines in the LEMMS energy range are power law fits, with associated spectral indices and uncertainties. Note that an error was made in the calculation of LEMMS differential intensities at the quasi-parallel shock that were reported by Masters et al. (2013), which are too high by a factor of 100. This has been corrected here, where quantitative comparison is relevant for our conclusions.

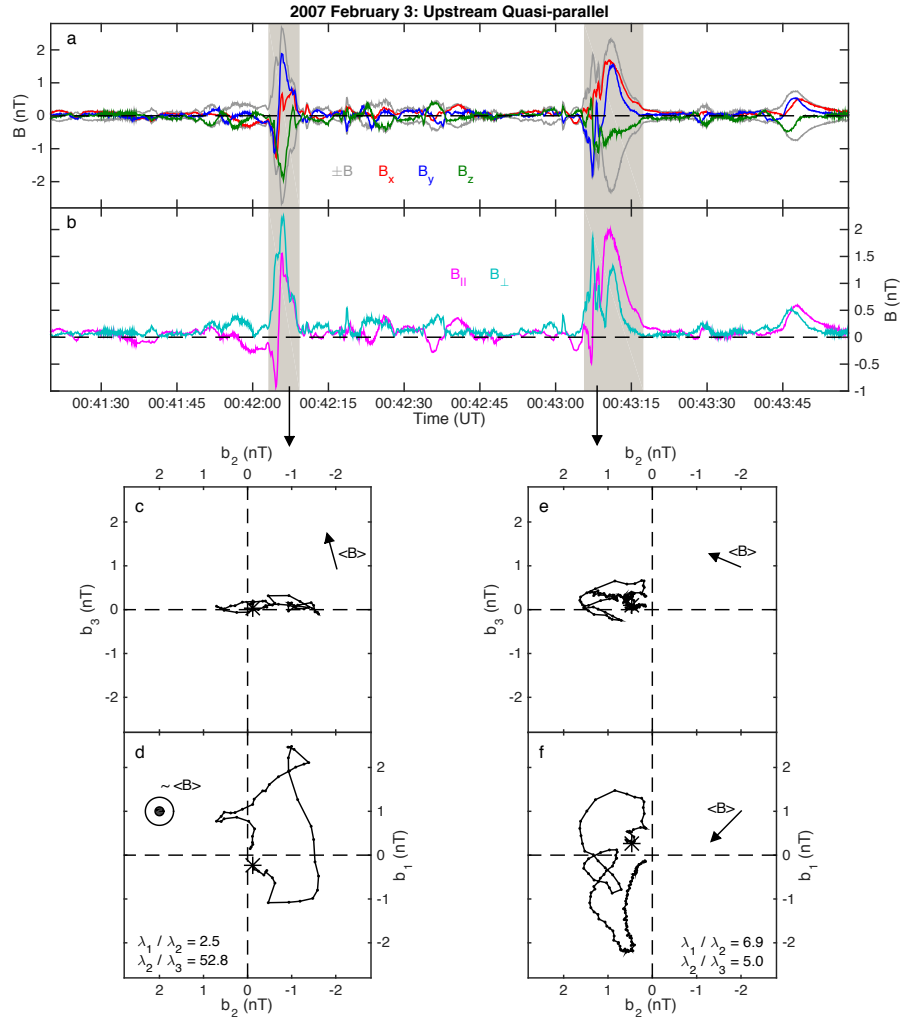

Figure 4. Magnetic field measurements made by Cassini on 2007 February 3 during an interval when the spacecraft was upstream of Saturn's quasi-parallel bow shock and two Short Large-Amplitude Magnetic Structures (SLAMS) were encountered. (a) Magnetic field magnitude and components in a Cartesian coordinate system. (b) Magnetic field components parallel and perpendicular to the background magnetic field (given as the average over the interval 00:10 to 01:00 UT). (c-f) Hodograms of the magnetic field measurements made during each structure, each shown in a coordinate system derived from minimum variance analysis. The maximum, intermediate, and minimum variance directions are $b_1$, $b_2$, and $b_3$, respectively. Ratios of associated eigenvalues are given ($\lambda_1$, $\lambda_2$, $\lambda_3$). The first field measurement in the time series is indicated by a star, and the projection of the background magnetic field vector, $<B>$, normalized to unity is given in each panel.

Acknowledgements

We thank Cassini instrument Principal Investigators S. M. Krimigis, D. T. Young, and J. H. Waite. This work was supported by UK STFC through consolidated grants to MSSL/UCL and Imperial College London. AM is supported by a Royal Society University Research Fellowship. LS was supported by Polish NSC grant 2016/22/E/ST9/00061. AHS is supported by NASA through Contract 1415150 with Jet Propulsion Laboratory.